\documentclass[reprint,superscriptaddress,aps,pre]{revtex4-1}

\usepackage{graphicx}
\usepackage{amssymb}
\usepackage{amsmath}
\usepackage{epsfig}
\usepackage{chemarr}
\usepackage{url}
\usepackage{natbib}
\usepackage{dcolumn}
\usepackage{bm}
\usepackage{color}

\def\be{\begin{equation}}
\def\ee{\end{equation}}
\def\bmu{\begin{multline}}
\def\bea{\begin{eqnarray}}
\def\eea{\end{eqnarray}}

\def\p{\partial} 

\def\f{\frac}

\def\l{\left(}
\def\r{\right)}

\begin{document}
\title{The self-tuned sensitivity of circadian clocks}

\author{Kabir Husain}
\affiliation{James Franck Institute and Department of Physics, University of Chicago, Chicago IL, USA}
\author{Weerapat Pittayakanchit}
\affiliation{James Franck Institute and Department of Physics, University of Chicago, Chicago IL, USA}
\author{Gopal Pattanayak}
\affiliation{Department of Molecular Genetics and Cell Biology, University of Chicago, Chicago IL, USA}
\author{Michael J Rust}
\affiliation{Department of Molecular Genetics and Cell Biology, University of Chicago, Chicago IL, USA}
\author{Arvind Murugan}
\email{amurugan@uchicago.edu}
\affiliation{James Franck Institute and Department of Physics, University of Chicago, Chicago IL, USA}
\begin{abstract}

Living organisms need to be sensitive to a changing environment while also ignoring uninformative environmental fluctuations. Here, we show that the circadian clock in \textit{Synechococcus elongatus} can naturally tune its environmental sensitivity, through a clock-metabolism coupling quantified in recent experiments. The metabolic coupling can detect mismatch between clock predictions and the day-night light cycle, and temporarily raise the clock’s sensitivity to light changes and thus entrain faster. We also analyze analogous behaviour in recent experiments on switching between slow and fast osmotic stress response pathways in yeast. In both cases, cells can raise their sensitivity to new external information in epochs of frequent challenging stress, much like a Kalman filter with adaptive gain in signal processing. Our work suggests a new class of experiments that probe the history-dependence of environmental sensitivity in biophysical sensing mechanisms.

\end{abstract}

\keywords{ }
\maketitle

Living organisms do not perceive their environment in an objective manner but often in the context of prior expectations or predictions of what the environment might be. Many examples of such prior expectations -- i.e., internal models of the external world --  are found in neuroscience \cite{Laughlin1981-hz}, but can also be found in metabolic dynamics of yeast \cite{Mitchell2015-oa} and bacteria \cite{Tu2008-dm}, the rhythms generated by free-running circadian clocks \cite{Winfree2001-pr}, receptor signalling cascades, and the immune system \cite{Mayer2015-dj}.

Combining predictions with measurements requires care, as both data might be unreliable. In the 1960s, Kalman \cite{Kalman1960} introduced a simple iterative scheme to optimally update predictions with measurements that has found applications from Apollo 11 \cite{Grewal2010-aw} to particle tracking in microscopy \cite{Wu2010} and synthetic genetic circuits in living cells \cite{Zechner2016-ig}. While the exact mathematics of Kalman filters is unlikely to apply to biology, the Bayesian idea at the heart of Kalman filtering is broadly applicable -- i.e. predictions must be updated by measurements using an iteratively computed weight that reflects their respective unreliabilities. However, it is not clear whether a Kalman-like adaptive sensitivity to new external information can be easily implemented at the cellular level. Indeed, unlike routine feedback regulation \cite{Huang2000-lb}, the quantity of physiological interest - e.g., osmotic pressure, circadian time - is not itself regulated in a Kalman strategy, but rather the rate at which that quantity is updated by new information.

Here, by analyzing two disparate systems, we argue that the ingredients needed for self-tuned sensitivity to new environmental information are readily found in biology. We first analyze recent quantitative experiments on the interaction between circadian clocks and metabolism in the photosynthetic cyanobacterium \textit{Synechococcus elongatus} \cite{Pattanayak2014-bv}. Here, the free running KaiABC-protein based circadian clock serves as an unreliable internal model of the external 24 hour day-night cycle of light on earth, `entrained' by periodic changes in external light. 

We interpret recent experiments to argue the sensitivity of the clock to light is tunable, since this sensitivity is controlled by the cell's metabolic state, in particular the availability of energy storage compounds such as glycogen. We further demonstrate that, since glycogen metabolism is controlled by the clock, the metabolically-coupled clock effectively tunes its own sensitivity, reaching values appropriate for different environmental conditions.

We then discuss similar behavior in stress response pathways in yeast. Recent experiments show how information from fast and slow osmolarity sensing pathways are combined to show the high speed of the fast pathway but retain the low error of the slow pathway\cite{Granados2017-fn}. We find that this behavior can be explained if the balance between these two pathways switches in a time-dependent manner.

We conclude with general results on when Kalman-like tunable sensitivity is biologically advantageous. We show that self-tuned sensitivity can break speed-accuracy (or gain-bandwidth) trade-offs in sufficiently heterogeneous environments, e.g., when the circadian clock switches between distinct epochs of high and low noise. Each distinct epoch needs to persist long enough to allow self-tuning mechanisms such as metabolic feedback or osmolarity mismatch to raise or lower the sensitivity as needed. Taken together, our results suggest new kinds of experiments that can reveal the phenotypic adaptation of sensitivity to new information in biophysical sensing.

\section{Mismatch sensing through metabolic coupling}

\begin{figure*}
\includegraphics[width=\textwidth]{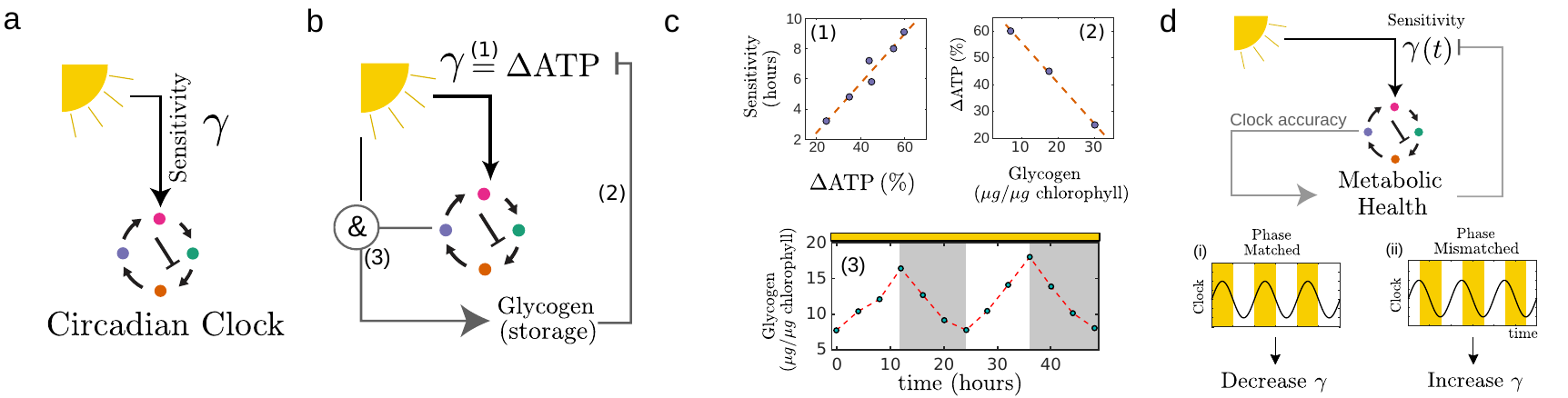}
\caption{ \label{fig:PRC} \textbf{The clock-metabolism coupling in \textit{S. elongatus} can self-regulate sensitivity to external light.}
\textbf{a} The light sensitivity $\gamma$ quantifies how quickly a free-running circadian clock is phase entrained by the external day-night light cycle. \textbf{b} Sensitivity $\gamma$ in \textit{S. elongatus} is self-regulated
by the clock-metabolism feedback with \textbf{c} experimentally quantified couplings \cite{Pattanayak2014-bv}. (1) Phase response curve height (a proxy for $\gamma$) grows with $\Delta$ATP = ATP$_{\text{day}}$- ATP$_{\text{night}}$. (2) $\Delta$ATP falls with increasing intracellular glycogen levels. (3) Glycogen production is gated by the clock; hence glycogen levels fall during subjective night (gray) in constant light. \textbf{d} Consequently, sensitivity $\gamma$ is dynamic, and can be tuned by clock accuracy, i.e. when clock output is mismatched with day-night light signals, glycogen falls and hence $\gamma$ increases.
}
\end{figure*}

\indent We consider free-running circadian clocks entrained to diurnal changes in light. Free running clocks show sustained periodic rhythms even in the absence of external periodic light or temperature signals and can be very complex, involving dozens of proteins and genes as in the case of the mammalian clock \cite{Leloup2003-rg}. 

\indent No matter how complex the clock, we can define an effective parameter -- `sensitivity' $\gamma$ -- that quantifies the coupling of the clock to an external entraining signal such as light. For example, $\gamma$ can be experimentally defined as the height of the `phase response curve', i.e., as the largest clock phase change in response to a single dark pulse administered at different times of the day \cite{Winfree2001-pr}. For simplicity, we consider light to be the only entraining signal for the clock.

\indent While the sensitivity is usually thought of as a fixed parameter, recent experiments on \textit{S. elongatus} have identified components that suggest a dependence on the recent history of clock performance. In particular, the sensitivity is set by a metabolic variable, glycogen, which itself is regulated by the history of clock accuracy.

\indent We quantify the link between clock and metabolism by analysing data from \cite{Pattanayak2014-bv}. Both \textit{in vivo} and \textit{in vitro} data suggest that the difference between day and night time ATP levels sets $\gamma$ (Figure \ref{fig:PRC}b); that is,

$$\gamma \propto \Delta \text{ATP} = \text{ATP}_{\text{day}} - \text{ATP}_{\text{night}} $$

\indent While day-time ATP levels are set by the rate of photosynthesis, and thus light intensity, night time ATP is produced from the cell's intracellular storage form of glucose, glycogen \cite{Pattanayak2014-bv}. Fig \ref{fig:PRC}c shows \textit{in vivo} data for the dependence of $\Delta$ATP on glycogen: increased glycogen levels increases $\text{ATP}_{\text{night}}$ and thus reduces sensitivity $\gamma$.

Critically, glycogen levels are in turn affected by  clock-environment mismatch. Data shown in Fig .\ref{fig:PRC}c from \textit{S. elongatus} \cite{Pattanayak2014-bv} grown in constant light shows that glycogen is produced only when it is both \textit{objectively} and \textit{subjectively} day, and degraded otherwise. We model these facts using,
\begin{equation} \label{eq:glycogendynamics}
    \tau_{\gamma} \frac{d \text{Gly}}{dt} = - \lambda \, \text{Gly} + \alpha \underbrace{ \Theta[\theta(t)] s(t)}_{\text{mismatch}}
\end{equation}
where $\Theta[\theta(t)] = 1$ if the clock state $\theta(t)$ corresponds to subjective day and $\Theta[\theta(t)] = 0$ otherwise and the external light $s(t)=1$ during the day and $s(t) = 0$ otherwise. Thus the production term is present only when it is objectively day ($s=1$) and also {subjectively} day ($\Theta(\theta) =1$). If the clock is out of phase with the external day-night signal, the hours of sunlight when the clock is in the night state are wasted in terms of glycogen production. Thereby, clock-environment mismatch raises the sensitivity $\gamma$, Fig. \ref{fig:PRC}d.

\begin{figure*}
\includegraphics[width=\textwidth]{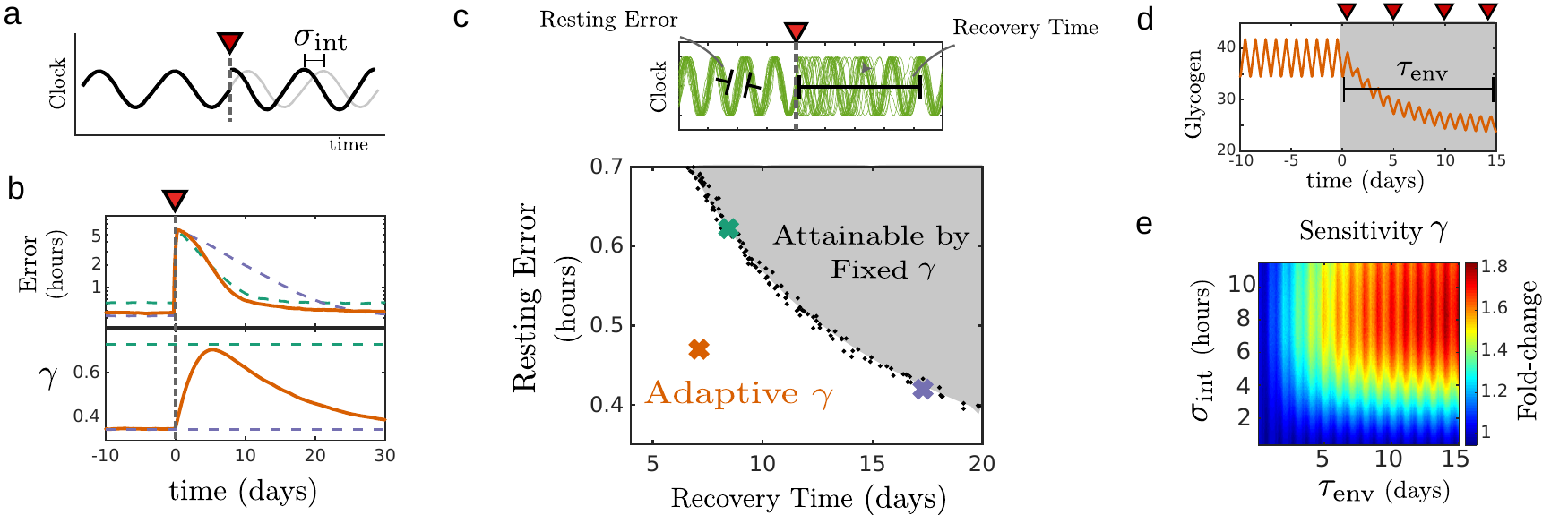}
\caption{ \label{fig:ClockGain} \textbf{Self-tuned sensitivity allows fast and yet accurate response in heterogeneous environments}.
\textbf{a} 
\textbf{b} Clock error (average error from objective time, top) and sensitivity $\gamma(t)$ (bottom) in response to a sudden phase shift (red triangle, a). By raising sensitivity $\gamma$ only when necessary, the self-tuned clock  (orange) entrains as fast as a fixed high $\gamma$ clock (green) but with the resting error of a low (purple) $\gamma$ clock, thus \textbf{c} beating speed-accuracy trade-offs inherent to fixed-$\gamma$ clocks (black dots).
\textbf{d} Fitting the data in Fig. \ref{fig:PRC}, we simulated an epoch of random repeated shifts of clock state by a typical amount $\sigma_{\text{int}}$ hours, causing repeated mismatch with environmental light. \textbf{d} Intracellular glycogen, averaged over $N=200$ cells, falls during this epoch; thus, \textbf{e} clock sensitivity (quantified by PRC height, after a period $\tau_{\text{env}}$ of disturbances, relative to sensitivity in undisturbed conditions) rises.
}
\end{figure*}

What are the benefits of self-tuned sensitivity in a circadian clock? We explored this in a minimal model of a generic circadian clock, consisting only of a phase oscillator $\theta(t)$ entrained by the external light signal $s(t)$. First, we characterised fixed sensitivity clocks subject to internal fluctuations, modeled by discrete events that shift the clock phase by an amount $\sigma_{\text{int}}$ (Fig \ref{fig:ClockGain}a). Such fluctuations can result from various forms of stress, such as periods of rapid cell division \cite{Teng2013-ax}. 

\indent As shown in Fig. \ref{fig:ClockGain}b,c, a large $\gamma$ re-entrains the clock to  external light quickly after a phase perturbation, but also rendering the clock sensitive to external fluctuations in light (say, due to weather patterns \cite{Troein2009-uz}). Conversely, a low $\gamma$ clock is robust against light fluctuations but is slow to entrain. The resultant trade-off is a manifestation of speed-accuracy trade-offs seen in such disparate fields as photoreceptor signal transduction \cite{Detwiler2000-wr}, neural decision making \cite{Heitz2014,Piet2018} , cellular concentration sensing\cite{Siggia2013-la,Govern2012-gm}, immunology \cite{Francois2016}, and control theory (e.g., the gain-bandwidth tradeoff \cite{Bechhoefer2005-iz}). We reasoned that a dynamic sensitivity could overcome this tradeoff.

\indent Inspired by the metabolic coupling in \textit{S. elongatus}, we augmented the model with a dynamic $\gamma(t)$ set by clock-environment mismatch: that is, $\gamma(t)$ is raised when clock phase $\theta(t)$ and the measured time of day $s(t)$ differ significantly (see SI). Simulations show that this mismatch feedback lets (Fig. \ref{fig:ClockGain}b)  $\gamma(t)$ idle at low sensitivity when well-entrained 
but transiently raises $\gamma$ to re-entrain the clock when needed. In this way, modulating sensitivity $\gamma$ by a memory of recent clock performance overcomes trade-offs inherent to fixed-$\gamma$ clocks (Fig. \ref{fig:ClockGain}c).

\indent To understand the conditions under which the metabolic feedback in \textit{S. elongatus} modulates clock gain, we fit the data in Fig. \ref{fig:PRC}c to construct a minimal model of the Kai oscillator with the measured glycogen feedback and dynamics (SI). We keep as a free parameter the decay rate $\lambda$ in Eq. \ref{eq:glycogendynamics}, whose value sets the resting glycogen level in well-entrained cells. Subjecting simulated cells to transient periods of high internal fluctuations lasting time $\tau_{\text{env}}$, we find that repeated phase shifts compromise glycogen storage, Fig \ref{fig:ClockGain}d. Measuring the phase response curve of our simulated cells, we find that the clock sensitivity correspondingly rises to significantly higher values ($\sim 2$ fold increase in PRC height), if these periods of repeated stress last long enough (large $\tau_{\text{env}}$) and are intense enough (large $\sigma_{\text{int}}$) so as to significantly change glycogen levels away from their resting values; see Fig \ref{fig:ClockGain}e. 

\indent Thus, while the phase response curve and sensitivity are usually thought of as fixed properties of a circadian clock \cite{Winfree2001-pr}, here we find that they can be tuned by the recent history of clock performance. Our framework generates testable predictions: while the perturbations in Fig \ref{fig:ClockGain} represent internal fluctuations, they could also represent `jet lag', i.e., jumps in the phase of an artificial light signal in the lab. In the Discussion, we describe experimental protocols to detect such history-dependent sensitivity. In either case, whether it be internal fluctuations or an irregular external signal, the organism would benefit from a higher sensitivity $\gamma$ and faster entrainment rates during such epochs.

\section{Self-tuned sensitivity in Osmolarity Response}

\begin{figure*}
\includegraphics[width=\textwidth]{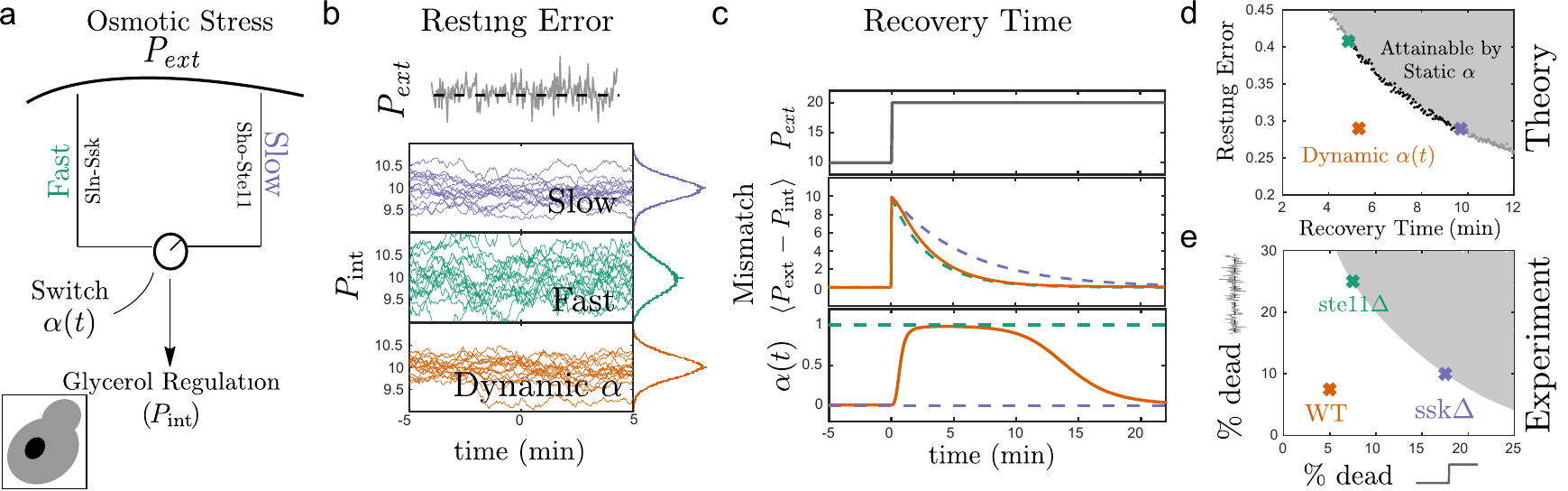}
\caption{ \textbf{Dynamic switching between fast and slow osmotic pressure response pathways in \textit{S. cerevisiae}.} 
\textbf{a} External osmotic pressure $P_{\text{ext}}$ affects internal glycerol production through a fast (high-$\gamma$) phosphorelay pathway and a slow (low-$\gamma$) kinase cascade. We combine the two pathways using a time-dependent relative weight $\alpha(t)$ set by osmotic mismatch $P_{\text{ext}} - P_{\text{int}}$.
\textbf{b} The dynamic switching $\alpha(t)$ (orange) model filters fluctuations in $P_{\text{ext}}$ as effectively as the slow (purple) pathway operating in isolation, but also \textbf{c} matches the fast pathway's speed in recovering from an osmotic shock.
\textbf{d} Hence the dynamic $\alpha(t)$ model beats the speed-accuracy trade-off inherent to any fixed-$\alpha$ circuit (black dots). \textbf{e} Experiments in \cite{Granados2017-fn} reveal a similar violation of the trade-off by wild type cells as compared to single pathway knockouts by measuring cell death in response to fluctuations and to step changes of $P_{\text{ext}}$.      \label{fig:yeast}}
\end{figure*}

Self-tuned sensitivity to new environmental information is broadly applicable beyond  metabolically-coupled clocks. Here we model recent experiments showing similarly tuned sensitivity in the osmolarity regulation pathway in the budding yeast, \textit{S. cerevisiae}. 

Sudden external changes in osmolyte concentration can lead to physical rupture of cells if not rapidly counteracted \cite{Hohmann2002}. \textit{S. cerevisiae} reacts to an osmotic shock by producing intracellular glycerol in response \cite{Saito2012}. Interestingly, the signaling between membrane receptors and glycerol production occurs via two distinct upstream branches that converge on the MAP kinase Hog1 in a Y-shaped motif. \cite{Posas1996,Posas1997,Hersen2008}. In isolation, one of the pathways -- the two-component Sln-SSk1 phospho-transfer -- leads to a fast but inaccurate response, while the other pathway -- the Sho-Ste11 kinase cascade -- is slow but accurate in restoring osmotic equilibrium \cite{Granados2017-fn}. Strikingly, the wild-type, which combines information from both pathways, manages to show the speed of the fast pathway but the error of the slow pathway \cite{Granados2017-fn}.

\indent Representing the signaling activity of each branch at time $t$ by $M_{\text{sln}}(t)$ and $M_{\text{sho}}(t)$, we model the joint regulation of glycerol as,
\be \label{eq:dynalpha}
\f{\p}{\p t} \text{gly} = \alpha(t) \, \gamma_{\text{sln}} M_{\text{sln}}(t) + (1-\alpha(t)) \, \gamma_{\text{sho}} M_{\text{sho}}(t)
\noindent 
\ee

where $\gamma_{\text{sln}}$ and $\gamma_{\text{sho}}$ are the response speeds of each pathway, with $\gamma_{\text{sho}} = \gamma_{\text{sln}}/2$ as in the experiments of \cite{Hersen2008}. Here, the weight factor $0 < \alpha <1$ prescribes the influence of each upstream pathway (Fig. \ref{fig:yeast}a). One can consider more complex non-linear models of joint regulation; our results below only depend on whether the relative importance of the two pathways is static or dynamic.
 
\indent Simulating the model, we reproduced the speed and accuracy behaviors of each branch in isolation by fixing $\alpha = 0$ and $1$ to emulate the Sln and Sho knockouts respectively; see Fig. \ref{fig:yeast}c, d. We then explored joint, but static regulation of glycerol: a fixed $\alpha(t) = \text{const.}$ leads to a trade-off between speed and accuracy, just as with either pathway in isolation and unlike the experiment \cite{Granados2017-fn}; see Fig.\ref{fig:yeast}d and e. In the SI we argue that this limit corresponds to a single upstream pathway with a fixed, effective response speed between $\gamma_{\text{sln}}$ and $\gamma_{\text{sho}}$.

\indent We can explain the breaking of the trade-off by the wild-type if the information from the two pathways is integrated instead with a dynamic weight $\alpha(t)$ (Fig. \ref{fig:yeast}c, d), that is raised by an osmotic pressure imbalance $\Delta P =P_{\text{ext}} - P_{\text{int}}$ (i.e., mismatch; Fig. \ref{fig:yeast}c) and kept low otherwise (Fig. \ref{fig:yeast}d). Thus, only by dynamically weighting inputs from each upstream pathway does the wild type leverage the desirable features of both.

This tunable speed of response in the yeast system is remniscent of the circadian clock presented above. Unlike with the clock-glycogen coupling, however, the exact molecular mechanism responsible for tuning $\alpha(t)$ is currently unknown. Our model regulates $\alpha(t)$ according to the mismatch $P_{ext}-P_{int}$. The model in \cite{Granados2017-fn} explains experimental data using mutual inhibition between the two arms; such inhibition effectively implements a time-varying factor $\alpha(t)$ as well. Independent of the detailed molecular mechanism, the experimental data of \cite{Granados2017-fn}, replotted in Fig.\ref{fig:yeast}e, on speed and error for the wild type compared to knockouts presents a convincing case of self-tuned sensitivity.

\section{Discussion}

\begin{figure*}[t]
\includegraphics[width=\linewidth]{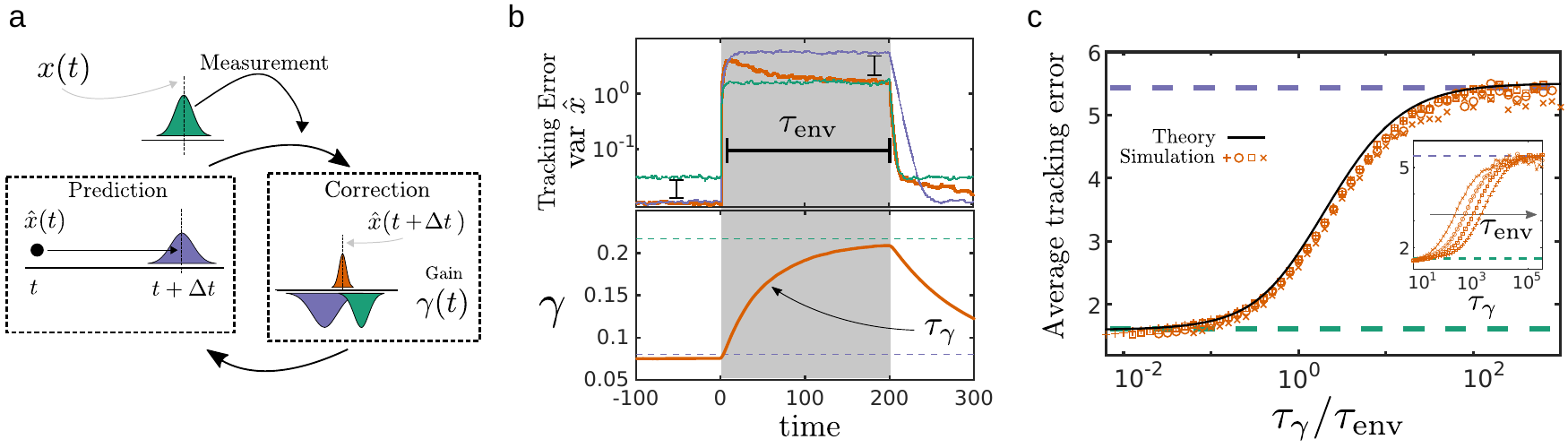}
\caption{\label{fig:kalman} \textbf{Adaptive gain out-performs fixed-gain in sufficiently heterogeneous environments.}
\textbf{a} A generic Kalman filter iteratively estimates the position $x(t)$ of a moving particle by correcting predictions (purple) with measurements (green) that are weighted by a finite gain $\gamma(t)$.
\textbf{b} (bottom) Adaptive $\gamma(t)$ (orange, Eq. \ref{eq:mismatch}) rises in response to a noisy environment (
gray box) lasting time $\tau_{\text{env}}$. (top) Resulting error in tracking. (purple, green, orange are fixed low, fixed high, adaptive $\gamma$ resp.). 
\textbf{c} Adaptive $\gamma$ lowers error relative to fixed low  $\gamma$ (purple) only when $\tau_{\text{env}} > \tau_{\gamma}$. We varied $\tau_{\text{env}}$ (symbols) and $\tau_{\gamma}$ and collapsed by scaling $\tau_{\gamma}/\tau_{\text{env}}$ (inset: uncollapsed). Solid black curve is an approximate analytical calculation (see SI). 
}
\end{figure*}

We have presented experimentally constrained quantitative models of two biological systems that navigate a trade-off between speed and accuracy by self-regulating their sensitivity. We now present a general framework for self-tuned sensitivity, based on Kalman filtering, that encompasses both systems. We use this simplified general framework to demonstrate what the important effective parameters are, on what timescales these self-tuned mechanisms are useful and what kinds of experiments can reveal them.

Kalman filtering is an iterative Bayesian approach to combine uncertain measurements of the environmental state with uncertain internal predictions (or expectations) of what the environmental state should be. Kalman filters are usually presented as a prediction-measurement-update cycle. For simplicity, consider a (discrete-time) Kalman filter for tracking a particle moving in one dimension with average velocity $v$ whose position is only periodically measured every $\delta t$ seconds. Between these measurements, we can estimate (or predict) the particle position to be $x_P(t) = \hat{x}(t- \delta t) + v \delta t$. Such predictions are assumed to be unreliable with variance $\sigma^2_{\text{int}}$, e.g., because of fluctuations in particle velocity. At the end of this $\delta t$ interval, the particle is measured to be at $x_{M}(t)$ with uncertainty $\sigma^2_{\text{ext}}$ relative to the real position. Since measurements and predictions are both unreliable, predictions must be corrected by this measurement with a finite sensitivity $\gamma$,
$$ \hat{x}(t) = (1-\gamma) \underbrace{x_P(t)}_{\text{Prediction}} + \; \gamma \underbrace{x_M(t)}_{\text{Measurement}}.$$
The process then repeats with the corrected estimate, $\hat{x}(t)$. 

Here, $\gamma$ reflects sensitivity to new external information; large $\gamma$ rapidly updates the internal state when internal and measured values disagree. Kalman's key idea was to iteratively update $\gamma$ over time so as to reflect the relative unreliabilities of measurements $x_M$ and internal predictions $x_P$. The literature contains numerous ways in which $\gamma$ can be updated over time. Motivated by our biological examples, we focus on feedback based on mismatch (also called a generalized or adaptive Kalman filter \cite{Rutan1991}),
\begin{equation}\label{eq:mismatch}
    \tau_{\gamma} \frac{d\gamma}{dt} =  \kappa \underbrace{\vert \hat{x}(t) - x_M(t) \vert}_{\text{mismatch}} - \gamma
\end{equation}
With this general simplified setup, we investigate when such self-tuned sensitivity can provide an advantage. 
We compute the average tracking error $\mbox{Var}(\hat{x}(t) - x(t))$ in a heterogeneous environment where predictions transiently have high error $\sigma_{\text{int}}$ for periods of length $\tau_{\text{env}}$.

As shown in Fig.\ref{fig:kalman}b, the adaptive strategy initially idles at low $\gamma$ but when predictions become noisy, $\gamma$ starts to rise towards a high value $\gamma_{\text{hi}}$, thus lowering error. However, if $\tau_{\text{env}}$ is too short, $\gamma$ cannot reach $\gamma_{\text{hi}}$ before the epoch ends. We find that the mismatch-mediated feedback is only effective when (see SI):
\be
\f{\tau_{\gamma}}{\tau_{\text{env}}} < \l 1 + \f{\kappa \, \sigma_{\text{int}}}{2\sqrt{\pi}\gamma^{3/2}_{\text{hi}}} \r
\ee
\indent Thus, only when $\tau_{\text{env}}$ is sufficiently long, and the stressful environment sufficiently adverse (high $\sigma_{\text{int}}$), does the adaptive Kalman filter leverage the benefits of a dynamic $\gamma$ in the noisy environment, as seen in Figure \ref{fig:kalman}(c).

The circadian clock and yeast stress response can be seen as examples of such generalized Kalman models. The clock corresponds to a model where $x$ is periodic and $v \sim 1/24$ hours. The epoch of high $\sigma_{\text{int}}$ could correspond to epochs of high internal fluctuations in clock phase (e.g., epochs of rapid growth \cite{Teng2013-ax}) that would benefit from fast and frequent re-entrainment. Osmolarity signaling corresponds to models with $v=0$, i.e., the internal model assumes osmolarity is not changing in order to reject high frequency fluctuations in external pressure. Here, epochs of frequent real changes in external osmotic pressure are mathematically captured by epochs of high $\sigma_{\text{int}}$ in the Kalman framework. Finally, the experiments in \cite{Pattanayak2014-bv} suggest $\tau_{\gamma}$ of several days for clocks, while experiments in \cite{Granados2017-fn} suggest a fast $\tau_{\gamma}$ for yeast that provides a benefit even for a single step change in osmolarity.

Our results here suggest how to design experiments to reveal self-tuned sensitivity mechanisms --- experiments need to measure sensitivity after a period of priming $\tau_{\text{env}}$ that lasts longer than the feedback timescale $\tau_{\gamma}$; further, the intensity of perturbations $\sigma_{\text{int}}$ during this interval need to be strong enough. 

In the context of the clock, experiments could, for example, measure the phase response curve after a period of priming. The priming protocol could use light-dark cycles, each an \emph{average} of 12 hours, but where night falls at an unexpected time, i.e., not at subjective dusk. The difference between subjective dusk and arrival of dark sets $\sigma_{int}$, while the total length of the protocol sets $\tau_{\text{env}}$. Our theory predicts that the measured sensitivity will be significantly greater after priming, if $\tau_{\text{env}} > \tau_{\gamma}$ and for large enough $\sigma_{\text{int}}$.

Feedback regulation is ubiquitous in biology. However, most known examples involve control or homeostasis problems where the quantity of physiological interest - e.g., osmotic pressure - is itself directly under feedback regulation; such regulation has been compared to PI controllers\cite{Huang2000-lb}. The Kalman-inspired feedback regulation of sensitivity discussed here is fundamentally distinct from such examples of control. Here, the \emph{sensitivity to new information} (often called gain) is under feedback regulation and the quantity of interest such as osmotic pressure is regulated based on such information. Further, our work shows how self-regulation of sensitivity can naturally arise from inevitable couplings in the cell - in \textit{S. elongatus}, the metabolic state is affected by clock performance and the metabolic state is, in turn, a globally relevant variable that affects clock sensitivity. 
We hope our work here will inspire experiments to test the history-dependence of sensitivity to new external information in diverse biophysical sensing pathways.

\textbf{Acknowledgements: } KH thanks the James S McDonnell Foundation for support via a Postdoctoral Fellowship. AM thanks the Simons Foundation for support. We are grateful to Amir Bitran, Ofer Kimchi, Mirna Kramar, Amanda Parker, and Ching-Hao Wang for early work on the project at the Cargese Summer School on Theoretical Biophysics (2017), to Catherine Triandafillou and Aaron Dinner for a careful reading of the manuscript, and to the Murugan and Rust groups for many critical discussions. We acknowledge the University of Chicago Research Computing Center for computing resources.

\bibliographystyle{unsrt}

\begin{thebibliography}{10}

\bibitem{Laughlin1981-hz}
S~Laughlin.
\newblock A simple coding procedure enhances a neuron's information capacity.
\newblock {\em Z. Naturforsch. C}, 36(9-10):910--912, September 1981.

\bibitem{Mitchell2015-oa}
Amir Mitchell, Ping Wei, and Wendell~A Lim.
\newblock Oscillatory stress stimulation uncovers an achilles' heel of the
  yeast {MAPK} signaling network.
\newblock {\em Science}, 350(6266):1379--1383, December 2015.

\bibitem{Tu2008-dm}
Yuhai Tu, Thomas~S Shimizu, and Howard~C Berg.
\newblock Modeling the chemotactic response of escherichia coli to time-varying
  stimuli.
\newblock {\em Proc. Natl. Acad. Sci. U. S. A.}, 105(39):14855--14860,
  September 2008.

\bibitem{Winfree2001-pr}
Arthur~T Winfree.
\newblock {\em The Geometry of Biological Time}.
\newblock Springer Science \& Business Media, June 2001.

\bibitem{Mayer2015-dj}
Andreas Mayer, Vijay Balasubramanian, Thierry Mora, and Aleksandra~M Walczak.
\newblock How a well-adapted immune system is organized.
\newblock {\em Proceedings of the National Academy of Sciences},
  112(19):5950--5955, May 2015.

\bibitem{Kalman1960}
R.~E. Kalman.
\newblock A new approach to linear filtering and prediction problems.
\newblock {\em Journal of Basic Engineering}, 82(1):35, 1960.

\bibitem{Grewal2010-aw}
M~S Grewal and A~P Andrews.
\newblock Applications of kalman filtering in aerospace 1960 to the present
  [historical perspectives].
\newblock {\em IEEE Control Syst.}, 30(3):69--78, June 2010.

\bibitem{Wu2010}
Pei-Hsun Wu, Ashutosh Agarwal, Henry Hess, Pramod~P. Khargonekar, and Yiider
  Tseng.
\newblock Analysis of video-based microscopic particle trajectories using
  kalman filtering.
\newblock {\em Biophysical Journal}, 98(12):2822--2830, jun 2010.

\bibitem{Zechner2016-ig}
Christoph Zechner, Georg Seelig, Marc Rullan, and Mustafa Khammash.
\newblock Molecular circuits for dynamic noise filtering.
\newblock {\em Proc. Natl. Acad. Sci. U. S. A.}, 113(17):4729--4734, April
  2016.

\bibitem{Huang2000-lb}
Y~Huang, M~I Simon, J~Doyle Proceedings~of the, and {2000}.
\newblock Robust perfect adaptation in bacterial chemotaxis through integral
  feedback control.
\newblock {\em National Acad Sciences}, 2000.

\bibitem{Pattanayak2014-bv}
Gopal~K Pattanayak, Connie Phong, and Michael~J Rust.
\newblock Rhythms in energy storage control the ability of the cyanobacterial
  circadian clock to reset.
\newblock {\em Curr. Biol.}, 24(16):1934--1938, August 2014.

\bibitem{Granados2017-fn}
Alejandro~A Granados, Matthew~M Crane, Luis~F Montano-Gutierrez, Reiko~J
  Tanaka, Margaritis Voliotis, and Peter~S Swain.
\newblock Distributing tasks via multiple input pathways increases cellular
  survival in stress.
\newblock {\em Elife}, 6, May 2017.

\bibitem{Leloup2003-rg}
Jean-Christophe Leloup and Albert Goldbeter.
\newblock Toward a detailed computational model for the mammalian circadian
  clock.
\newblock {\em Proc. Natl. Acad. Sci. U. S. A.}, 100(12):7051--7056, June 2003.

\bibitem{Teng2013-ax}
Shu-Wen Teng, Shankar Mukherji, Jeffrey~R Moffitt, Sophie de~Buyl, and Erin~K
  O'Shea.
\newblock Robust circadian oscillations in growing cyanobacteria require
  transcriptional feedback.
\newblock {\em Science}, 340(6133):737--740, May 2013.

\bibitem{Troein2009-uz}
C~Troein, Jcw Locke, M~S Turner, and A~J Millar.
\newblock Weather and seasons together demand complex biological clocks.
\newblock {\em Curr. Biol.}, 19(22):1961--1964, January 2009.

\bibitem{Detwiler2000-wr}
P~B Detwiler, S~Ramanathan, A~Sengupta, and B~I Shraiman.
\newblock Engineering aspects of enzymatic signal transduction: photoreceptors
  in the retina.
\newblock {\em Biophys. J.}, 79(6):2801--2817, December 2000.

\bibitem{Heitz2014}
Richard~P. Heitz.
\newblock The speed-accuracy tradeoff: history, physiology, methodology, and
  behavior.
\newblock {\em Frontiers in Neuroscience}, 8, jun 2014.

\bibitem{Piet2018}
Alex~T. Piet, Ahmed~El Hady, and Carlos~D. Brody.
\newblock Rats adopt the optimal timescale for evidence integration in a
  dynamic environment.
\newblock {\em Nature Communications}, 9(1), oct 2018.

\bibitem{Siggia2013-la}
Eric~D Siggia and Massimo Vergassola.
\newblock Decisions on the fly in cellular sensory systems.
\newblock {\em Proceedings of the National Academy of Sciences},
  110(39):E3704--12, September 2013.

\bibitem{Govern2012-gm}
Christopher~C Govern and Pieter~Rein ten Wolde.
\newblock Fundamental limits on sensing chemical concentrations with linear
  biochemical networks.
\newblock {\em Phys. Rev. Lett.}, 109(21):218103, November 2012.

\bibitem{Francois2016}
Paul Fran{\c{c}}ois and Gr{\'{e}}goire Altan-Bonnet.
\newblock The case for absolute ligand discrimination: Modeling information
  processing and decision by immune t cells.
\newblock {\em Journal of Statistical Physics}, 162(5):1130--1152, jan 2016.

\bibitem{Bechhoefer2005-iz}
J~Bechhoefer.
\newblock Feedback for physicists: A tutorial essay on control.
\newblock {\em Rev. Mod. Phys.}, January 2005.

\bibitem{Hohmann2002}
S.~Hohmann.
\newblock Osmotic stress signaling and osmoadaptation in yeasts.
\newblock {\em Microbiology and Molecular Biology Reviews}, 66(2):300--372, jun
  2002.

\bibitem{Saito2012}
H.~Saito and F.~Posas.
\newblock Response to hyperosmotic stress.
\newblock {\em Genetics}, 192(2):289--318, oct 2012.

\bibitem{Posas1996}
Francesc Posas, Susannah~M Wurgler-Murphy, Tatsuya Maeda, Elizabeth~A Witten,
  Tran~Cam Thai, and Haruo Saito.
\newblock Yeast {HOG}1 {MAP} kinase cascade is regulated by a multistep
  phosphorelay mechanism in the {SLN}1{\textendash}{YPD}1{\textendash}{SSK}1
  {\textquotedblleft}two-component{\textquotedblright} osmosensor.
\newblock {\em Cell}, 86(6):865--875, sep 1996.

\bibitem{Posas1997}
F.~Posas.
\newblock Osmotic activation of the {HOG} {MAPK} pathway via ste11p {MAPKKK}:
  Scaffold role of pbs2p {MAPKK}.
\newblock {\em Science}, 276(5319):1702--1705, jun 1997.

\bibitem{Hersen2008}
P.~Hersen, M.~N. McClean, L.~Mahadevan, and S.~Ramanathan.
\newblock Signal processing by the {HOG} {MAP} kinase pathway.
\newblock {\em Proceedings of the National Academy of Sciences},
  105(20):7165--7170, may 2008.

\bibitem{Rutan1991}
Sarah~C. Rutan.
\newblock Adaptive kalman filtering.
\newblock {\em Analytical Chemistry}, 63(22):1103A--1109A, nov 1991.

\end{thebibliography}

\end{document}


\setarrowdefault{,1.2,}
\setcompoundsep{3em}

\title{ {\Large The self-tuned sensitivity of circadian clocks} \\ {\normalsize SI} }

\author{K. Husain, W. Pittayakanchit, G. Pattanayak, M.J. Rust, A. Murugan}

\maketitle 

\section{Adaptive Gain in Circadian Oscillators}

\textit{Pertaining to Section I and Fig. 2 of the main text.}

\vspace{.2in}

\indent For pedagogical reasons, we begin our analysis with a generic model of a driven phase oscillator augmented with an adaptive gain circuit. The underlying equation of motion for the oscillator phase $\theta (t)$ is:

\be
\f{\p}{\p t}\theta(t) = \omega_0 + \gamma \, \cos\theta \, s(t)
\ee

\indent Here, $\omega_0$ is the intrinsic frequency of the oscillator. We measure time in units of days, and set $\omega_0 = 2\pi$. The parameter $\gamma$, as discussed in the main text, is the gain or, alternatively, the magnitude of the infinitesmal phase response curve whose shape is $\cos\theta$. The external signal $s(t)$ is decomposed into a regular signal and noise: $s(t) = s_0(t) + \eta(t)$. We take the regular signal $s_0(t)$ to be sinusoidal with frequency $\omega_0$: $s_0(t) = \sin \omega_0 t$, and the corrupting noise signal to be white with variance $\sigma_{\text{ext}}^2$:

\bea
 \langle \eta (t) \rangle &=& 0 \nn \\
 \langle \eta(t)\eta(t^{\prime}) \rangle &=& \sigma_{\text{ext}}^2 \, \delta(t-t^{\prime})
\eea

\indent We implement an adaptive gain by the following dynamics for $\gamma$:

\be
\tau_{\gamma} \, \f{\p}{\p t}\gamma = -\l \gamma - \gamma_0 \r + K_{\text{mismatch}} \underbrace{\l \f{1}{2} - s(t)\,\sin\theta \r}_{\mathcal{M}(\theta,s(t))}
\ee

\indent The parameter $K_{\text{mismatch}}$ quantifies the influence of the mismatch circuit on the gain dynamics. The form of the mismatch term $\mathcal{M}(\theta,s(t))$ is chosen such that, for sufficiently long $\tau_{\gamma}$, $\int_0^{\tau_{\gamma}} dt \, \mathcal{M}(\theta,s(t))$ evaluates to $0$ when $s(t)$ and $\theta(t)$ are in-phase, and $>0$ otherwise. $\gamma_0$ is the resting value of $\gamma$ in the absence of mismatch feedback. For the panels presented in Figure 2 in the main text, we use the parameters: $\sigma_{\text{ext}} = 0.5$, $\gamma_0 = 0.3$, $\tau_{\gamma} = 10$ days, and $K_{\text{mismatch}} = 3$. The equations are solved by Euler's method with a fixed time step $\text{d}t = 10^{-2}$.

\indent The response time and error of the clock is assessed in the following simulation. $N$ phase oscillators are allowed to entrain to the signal $s(t)$, each being exposed to an independent realisation of the external noise $\eta(t)$. The resting error of the clock is quantified by the difference between the internal and external time. Denoting by $\phi(t) = \omega_0 t$ the phase of $s(t)$, this is written:

\be
\text{Error:} \, \, \langle \, \arccos \l \cos \theta \cos \phi + \sin \theta \sin \phi \r \, \rangle
\ee

\noindent where the average is over the ensemble of $N$ phase oscillators. The resultant error in radians is converted to hours.

\indent The recovery time of the phase oscillator is measured as follows. An initially entrained population of oscillators is perturbed at $t = 0$ by a phase shift $\Delta \theta$, chosen to be a uniformly distributed value in $\left[-\phi_s, \phi_s \right]$. For the simulations presented in the main text, $\phi_s = 12$ hours. The system is then evolved under the external signal $s(t)$ until the population variability $\epsilon$, defined as $\epsilon = 1 - \sqrt{\langle \cos \theta \rangle^2 + \langle \sin \theta \rangle^2}$, falls to within $10 \%$ of its resting value (indicating that the population is once again entrained).

\section{Metabolic Feedback in the \textit{S. elongatus} Clock}

\textit{Pertaining to Section I and Fig. 1 of the main text.}

\vspace{.2in}

\subsection{Model}

\indent To mimic the limit cycle clock of \textit{S. elongatus}, we consider a Stuart-Landau oscillator of radius $R$ around $\vec{r}_0 = (x_0,y_0)$:

\bea \label{eq:KaiModel}
&& \f{\p}{\p t}x = -\omega_0 \, (y - y_0) + \alpha \l 1 - \f{(x-x_0)^2 + (y-y_0)^2}{R^2} \r (x - x_0) \nn \\ 
&& \f{\p}{\p t}y = \omega_0 \, (x - x_0) + \alpha \l 1 - \f{(x-x_0)^2 + (y-y_0)^2}{R^2} \r (y - y_0)
\eea

\indent When $\alpha >0$, and $\vec{r}_0$ is constant, the system settles into a circular limit cycle with time period $2\pi/\omega_0$. Once again, we measure time in units of days and set $\omega_0 = 2\pi$. We fix the parameter $\alpha = 10$.

\indent The oscillator couples to the external signal $s(t)$ through $\vec{r}_0(t)$. We choose coordinates by setting $y_0 = 0$ and:

\be
x_0(t) = \gamma \, (1 - s(t))
\ee

\indent The external signal $s(t)$ is taken to be a square wave (a `12-12 LD' cycle), with value $1$ during the day and $0$ during the night. Under these conditions, an entrained oscillator has $y >0$ during the day and $y <0$ during the night.

\indent In \textit{S. elongatus}, the Kai oscillator does not sense light directly but instead through metabolic intermediaries. We therefore take gain $\gamma$ to be a function of the difference between day and night intracellular ATP levels:

\be \label{eq:gammaATP}
\gamma = \gamma (\Delta \text{ATP})
\ee 

\indent The level of ATP during the day is fixed by photosynthesis, and therefore by light levels; however, ATP levels at night come from intracellular energy storage in glycogen. We therefore have:

\be \label{eq:ATPglyco}
\Delta \text{ATP} = \Delta \text{ATP} (\text{glycogen}) 
\ee 

\indent Finally, glycogen itself is a dynamic quantity: made during the day and consumed during the night. As we describe in the main text, previous work has suggested that glycogen production occurs only when it is both subjectively and objectively day (i.e., $s(t) = 1$ and $y > 0$). We therefore write for its dynamics:

\bea \label{eq:glycogenModel}
&& \f{\p}{\p t}\text{glycogen} = \lambda_g \alpha(t) - (1-\alpha(t)) \, k_g \times \text{glycogen} \nn \\
&& \alpha(t) = \Theta \l s(t)\times y(t) \r
\eea

\indent Here, $\lambda_g$ is the production rate of glycogen during the day, and $k_g$ is the degradation rate of glycogen during the night. The indicator variable $\alpha(t)$ is $1$ when both $s(t)$ and $y(t)$ are positive, and $0$ otherwise.

\begin{figure*}
\includegraphics[width=\linewidth]{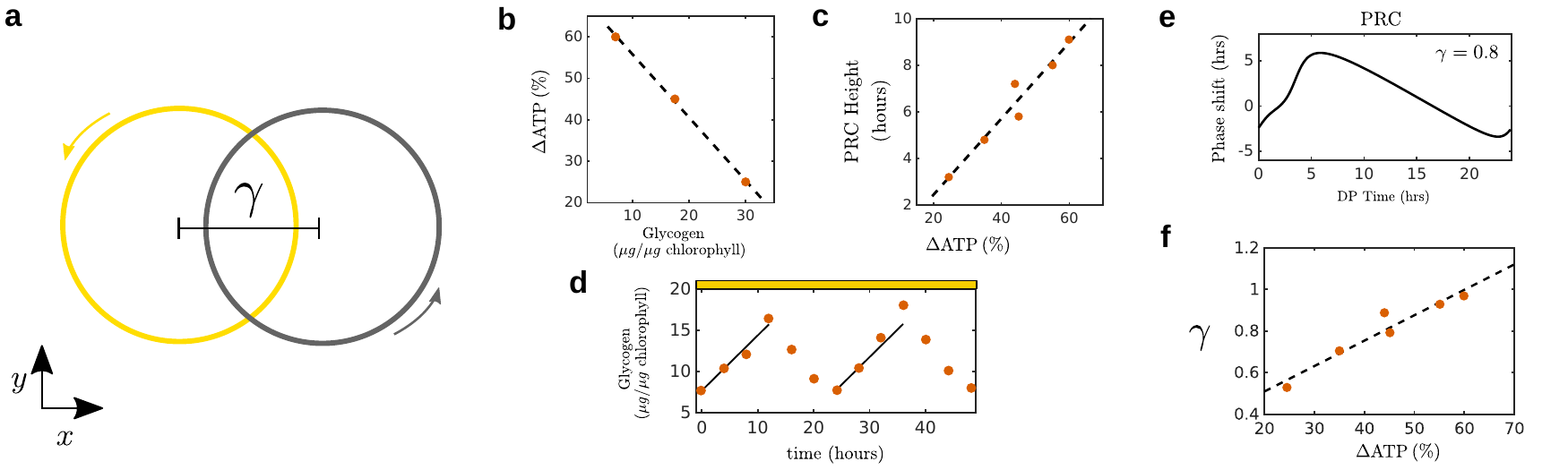}
\caption{ \label{fig:SI}
\textbf{a} The phase portrait of Eq. \ref{eq:KaiModel}, showing the $L$ (day, yellow) and $D$ (night, grey) limit cycles. The sensitivity $\gamma$, in this model, is defined as the distance between the two limit cycles. \textbf{b,c,d} Experimental data from \cite{Pattanayak2014-bv}; in \textbf{b}, dashed line shows Eq. \ref{eq:deltATPglycfit}, while in \textbf{d} the solid lines are predicted glycogen accumulation from Eq. \ref{eq:glycogenModel}. \textbf{e} A simulated PRC from Eqs. \ref{eq:KaiModel}, with a fixed $\gamma$ (here taken to be $0.8$). The height of the simulated PRC is measured for varying $\gamma$ and compared against the data in \textbf{c} to obtain \textbf{f}. The dashed line in \textbf{f} is Eq. \ref{eq:deltATPglycfit}.
}
\end{figure*}

\subsection{Simulated Noisy Epochs in the \textit{S. elongatus} clock}

\textit{Pertaining to Fig. 2 of the main text.}

\indent Eqs. \ref{eq:KaiModel} and \ref{eq:glycogenModel}, along with the algebraic relations Eqs. \ref{eq:gammaATP} and \ref{eq:ATPglyco}, constitute our model for the \textit{S. elongatus} circadian clock. To fix parameters and functional forms, we turn to experimental data. The data we have at hand are:

\begin{itemize}
	\item $\Delta \text{ATP}$ as a function of glycogen, Fig \ref{fig:SI}b.
	\item Phase response curve (PRC) height as a function of $\Delta \text{ATP}$, Fig \ref{fig:SI}c.
	\item Time-series of glycogen over one day, Fig \ref{fig:SI}d.
\end{itemize}

\indent First, we directly extract the functional form of $\Delta \text{ATP} (\text{glycogen})$, finding:

\be \label{eq:deltATPglycfit}
\Delta \text{ATP} = 72\% - 1.6\% \times \text{glycogen}
\ee

\noindent where $\Delta \text{ATP}$ is measured in percentage and glycogen in $\mu g$ per $\mu g$ chlorophyll.

\indent To extract the relationship Eq. \ref{eq:gammaATP}, we need to know how PRC height varies as a function of $\gamma$ in our model. We therefore perform simulated PRC `experiments' with Eq. \ref{eq:KaiModel}, mimicing the experimental protocol (i.e., LL interrupted by a five hour dark pulse \cite{Pattanayak2014-bv}).

\indent Matching values between the simulated PRC heights and the experimentally measured ones in Fig. \ref{fig:SI}, we compute a corresponding value of $\gamma$ for each $\Delta \text{ATP}$. We find that the relationship is roughly linear, and can be captured by:

\be \label{eq:gammaDeltATPfit}
\gamma = 0.26 + \l 1.2 \times 10^{-2} \r \times \Delta \text{ATP}
\ee

\indent Finally, we fix the dynamics of glycogen by measuring the gain in glycogen levels over a single day. A linear fit gives us $\lambda_g \approx 16$. We estimate the degradation rate by choosing one such that the average level of glycogen for a healthy, entrained cell is $\sim 40$ $\mu g$ per $\mu g$ chlorophyll, i.e., near the upper range probed experimentally (see Fig. \ref{fig:SI}b); we choose $k_g \approx 0.4$.


\indent With parameters fixed, we go on to expose our simulated cells to periods of stress. We use the Euler method to solve the equations of motion, with a time step $\text{d}t = 10^{-3}$. $N=2000$ cells are first entrained to a clean 12-12 LD signal $s(t)$. Then, at $t = 0$, each cell is exposed to a stochastic protocol of stress events. Each stress event corresponds to a phase shift of magnitude $\Delta \phi$ hours, uniformly distributed in the interval $\Delta \phi \in \left[ -\sigma_{int},\sigma_{int} \right]$. An exponentially distributed waiting time of mean value $1$ day separates stress events; the protocol lasts for time $\tau_{\text{env}}$. The external signal $s(t)$ continues as a 12-12 square wave during this time.

\indent At the end of the stress period, the cells are shifted to LL (i.e., the signal $s(t) = 1$). Each cell undergoes a standard PRC protocol (i.e., a 5 hour dark pulse) as described above, and the PRC height is normalised to the resting PRC height (i.e., that computed from a cell that has not undergone the stress procedure).

\section{The Osmotic Circuit}

\textit{Pertaining to Section II and Fig. 3 of the main text.}

\vspace{.2in}

\indent To model the regulation of glycerol, $g(t)$, by the external osmotic pressure $P_{\text{ext}}$, we first consider the simplest possible regulatory motif:

\be
\f{\p}{\p t} g(t) \, = -\f{1}{\tau} \l g(t) \, - P_{\text{ext}}(t) \r 
\ee

\indent Here, the timescale $\tau$ is the response time of the pathway, with the response speed $\gamma$ generically defined to be $1/\tau$. Solving,

\be
g(t) = \int dt^{\prime} \,\gamma \me^{-\gamma \l t - t^{\prime}\r} \, P_{\text{ext}}(t^{\prime})
\ee

\indent A large $\gamma$ corresponds to a rapid tracking of $P_{\text{ext}}(t)$ by $g(t)$. However, if the external signal contains noise, $P_{\text{ext}}(t) \to P_{\text{ext}}(t) + \eta(t)$, a large $\gamma$ (small $\tau$) is less able to average out the noise than a small $\gamma$ circuit \cite{Detwiler2000-wr}.

\indent As discussed in the main text, glycerol receives inputs from both the fast and slow signalling pathways; the simplest model consistent with this is:

\be
\f{\p}{\p t}g(t) = \underbrace{-\alpha(t)\,\f{1}{\tau_f} \l g(t) - P_{\text{ext}}(t) \r}_{\text{Slow pathway}} - \overbrace{ (1-\alpha(t))\,\f{1}{\tau_s} \l g(t) - P_{\text{ext}}(t) \r}^{\text{Fast pathway}} 
\ee

\indent Here, $\tau_f$ is the response time of the fast pathway and $\tau_s$ the response time of the slow pathway. From experiments in \cite{Hersen2008}, we estimate these to be $3$s and $6$s, respectively. The indicator variable $\alpha(t)$, which takes values between $0$ and $1$, decides which pathway regulates glycerol production. Finally, the external pressure is separated into a slowly changing component and rapid fluctuations, $P_{\text{ext}}(t) = P_0(t) + \eta(t)$. For these simulations, we take the noise $\eta(t)$ to be white, with variance $\sigma_{\text{ext}}^2 = 1$.

\indent In Figure 3 in the main text, we contrast a static $\alpha$ (i.e., $\alpha(t) =$ const) circuit against a dynamic $\alpha(t)$ one. The former is equivalent to a single pathway regulating glycerol production, with an effective response timescale between $\tau_f$ and $\tau_s$. The latter is implemented by writing $\alpha$ in terms of an auxiliary dynamical variable $\beta$, which measures the mismatch between internal and external pressure:

\bea
&& \alpha(t) = \f{\beta^n}{K_{\beta}^n + \beta^n} \nn \\
&& \f{\p}{\p t}\beta(t) = -\f{1}{\tau_{\gamma}} \l \beta(t) - \underbrace{\l P_{ext}(t) - g(t) \r}_{\text{Mismatch}} \r
\eea 

\indent The form of $\alpha$'s dependence on $\beta$ is chosen to resemble a switch, such that for much of the time $\alpha \approx 0$ or $1$; we choose the Hill co-efficient to be $n=4$. The other parameters are set as $K_{\beta} = 1$ and $\tau_{\gamma} = 6$s.

\indent To measure resting error and response time, we simulate the response of $N = 3000$ cells (with a timestep of $\text{d}t = 0.01$s) to a jump in external pressure from $P_0(t) = 10$ to $P_0(t) = 20$. The recovery time is measured as the average time taken for glycerol to come within $20 \%$ of its new value. The resting error is quantified as the population variance in glycerol levels just prior to the pressure jump.

\section{Adaptive Kalman Filter}

\textit{Pertaining to Section III and Fig. 4 of the main text.}

\vspace{.2in}

\indent We implement a discrete time Kalman filter for a particle with constant velocity $v$ as:

\bea \label{eq:vanillaKalman}
&& \text{Prediction: } x^{(P)}_t = \hat{x}_{t-1} + v \Delta t + \eta_{\text{int}}(t) \nn \\
&& \text{Measurement: } x^{(M)}_t = x_t + \eta_{\text{ext}}(t) \nn \\
&& \text{Update: } \hat{x}_{t} = x^{(P)}_t + \gamma_t \, \l x^{(M)}_t - x^{(P)}_t  \r
\eea

\indent Here, $x_t$ is the actual position of the particle at time $t$, $x^{(M)}_t$ is the measured position and $x^{(P)}_t$ is the predicted position. These are combined with the gain $\gamma_t$ to obtain the best estimate $\hat{x}_t$. $\eta_{\text{int}}$ and $\eta_{\text{ext}}$ represent the error in prediction and measurement, respectively; here, we take them to be normally distributed with mean $0$ and variances $\sigma_{\text{int}}^2$ and $\sigma^2_\text{ext}$

\indent We prescribe for the gain $\gamma_t$ the following dynamics:

\be \label{eq:vanillaKalmanwithChocolateChips}
\gamma_{t+1} = \gamma_{t} + \frac{\Delta t}{\tau_{\gamma}} \l \kappa \, \underbrace{ \vert \hat{x}_t - x^{(M)}_t \vert }_{\text{Mismatch }\mathcal{M}}- \gamma_t \r
\ee

\indent In the limit of large $\tau_{\gamma}$ this is approximated by the differential equation:

\be \label{eq:mismatchDynamics}
\tau_{\gamma} \frac{d\gamma}{dt} =  \kappa \vert \hat{x}_t - x^{(M)}_t \vert - \gamma
\ee

\noindent which is shown in the main text.

\indent We propagate the Kalman filter, Eqs. \ref{eq:vanillaKalman} and \ref{eq:vanillaKalmanwithChocolateChips}, numerically, with parameters $v = 0.1$, $\Delta t = 1$, $\kappa = 0.2$ and $\sigma_{\text{ext}} = 0.5$. The filter is first equilibrated in a `clean' environment with low internal noise, $\sigma_{\text{int}} = 0.01$; then, $\sigma_{\text{int}}$ is raised to a high value, $\sigma_{\text{int}} = 1$, for a time $\tau_{\text{env}}$.

\indent Defining the instantaneous error $\sigma \equiv \langle \l \hat{x}_t - x_t \r^2 \rangle$, we vary $\tau_{\gamma}$ and $\tau_{\text{env}}$ and compute the time-averaged error:

\be \label{eq:trackingError}
\text{Tracking error: } \f{1}{\tau_{\text{env}}} \int_0^{\tau_{\text{env}}} dt \, \sigma^2(t) \equiv \f{1}{\tau_{\text{env}}} \int_0^{\tau_{\text{env}}} dt \, \langle \l \hat{x}_t - x_t \r^2 \rangle
\ee

\indent For each value of $\tau_{\gamma}$ and $\tau_{\text{env}}$, the average is taken over an ensemble of $N = 600$ replicates. The resultant error is plotted in Figure 4c in the main text; we see that $\tau_{\text{env}}$ needs to be long enough for the mismatch-mediated feedback to raise the gain $\gamma$ and thereby lower the error. 

\indent To gain a more analytic understanding, we compute Eq. \ref{eq:trackingError} from an approximate solution of Eqs. \ref{eq:vanillaKalman} and \ref{eq:vanillaKalmanwithChocolateChips}. The coupled dynamics of $\hat{x}$ and $\gamma$ are too difficult to solve directly; we instead work in an `adiabatic' approximation -- valid for large $\tau_{\gamma}$ -- in which the statistics of $\hat{x}$ are always at steady state. That is, at time $t$ and gain $\gamma(t)$, the instantaneous variance of $\hat{x}$ is, from the steady state of Eq. \ref{eq:vanillaKalman}:

\be \label{eq:trackingSigma}
\sigma^2 (t) = \sigma^2 \l \gamma(t) \r \equiv \text{var }\hat{x} = \gamma^2 \sigma^2_{\text{ext}} + \f{(\gamma-1)^2}{1-(\gamma-1)^2} \sigma^2_{\text{int}}
\ee

\indent To solve for $\gamma(t)$, we average the mismatch term in Eq. \ref{eq:mismatchDynamics} to obtain:

\be \label{eq:adiabaticGamma}
\tau_{\gamma}\f{\p}{\p t}\gamma = -\gamma + \kappa \sqrt{\f{2}{\pi} \l \l \gamma^2 + 1 \r\sigma_{\text{ext}}^2 + \f{(\gamma-1)^2}{1-(\gamma-1)^2}\sigma_{\text{int}}^2 \r }
\ee

\indent For any particular values of $\sigma_{\text{ext}}$ and $\sigma_{\text{int}}$, the steady state value of $\gamma$ follows by setting the LHS to $0$. In our numerical protocol above, we change the value of $\sigma_{\text{int}}$ over time. Let $\gamma_{\text{lo}}$ be the steady-state value of $\gamma$ when $\sigma_{\text{int}}$ is low, and $\gamma_{\text{hi}}$ be the steady-state value of $\gamma$ when $\sigma_{\text{int}}$ is high. At $t = 0$ the noise statistics change from a low to a high $\sigma_{\text{int}}$. The relaxation of $\gamma$ from $\gamma_{\text{lo}}$ to $\gamma_{hi}$ may be approximately computed by linearising Eq. \ref{eq:adiabaticGamma} about $\gamma_{\text{hi}}$:

\bea \label{eq:approxAdiabaticGamma}
\f{\p}{\p t}\gamma &\approx& -\f{1}{\tau}\l \gamma - \gamma_{\text{hi}} \r \nn \\
\tau &=& \tau_{\gamma} \l1+ \kappa\sqrt{\f{\pi}{2}} \, \f{ (2-\gamma_{\text{hi}})^2\gamma_{\text{hi}}^3\sigma_{\text{ext}}^2 + (\gamma_{\text{hi}}-1)\sigma_{\text{int}}^2 }{(2-\gamma_{\text{hi}})^2\gamma_{\text{hi}}^2 \sqrt{ (1+\gamma_{\text{hi}}^2)\sigma_{\text{ext}}^2 - \f{(1-\gamma_{\text{hi}})^2\sigma_{\text{int}}^2}{(\gamma_{\text{hi}} - 2)\gamma_{\text{hi}}} }} \r^{-1}  \nn \\
&\approx & \tau_{\gamma} \, \l 1 + \f{\kappa \, \sigma_{\text{int}}}{2\sqrt{\pi} \gamma_{\text{hi}}^{3/2}} \r^{-1}
\eea

\noindent where, in the last line, we have expanded around small $\gamma_{\text{hi}}$. Only when $\tau_{\text{env}} > \tau$, the relaxation time, does $\gamma(t)$ reach $\gamma_{\text{hi}}$ and the error fall, as reported in the main text.

\indent Solving Eq. \ref{eq:approxAdiabaticGamma}, we obtain:

\be
\gamma(t) \approx \gamma_{\text{lo}} + \l \gamma_{\text{hi}} - \gamma_{\text{lo}} \r \exp(-t/\tau)
\ee

\indent Inserting this expression into Eq. \ref{eq:trackingSigma} and integrating from $t = 0$ to $t = \tau_{\text{env}}$, we obtain our (somewhat cumbersome) prediction for the average tracking error, Eq. \ref{eq:trackingError}:

\bea \label{eq:somethingofamouthful}
&& \f{1}{\tau_{\text{env}}}\int_0^{\tau_{\text{env}}} dt \, \sigma(\gamma(t)) = \f{1}{2} \l A \, \sigma_{\text{ext}}^2 + B \, \sigma_{\text{int}}^2 \r \nn \\
&& A = \f{\tau}{\tau_{\text{env}}}\l -3 \gamma_{\text{lo}}^2 + 4 \me^{-\tau_{\text{env}}/\tau} \gamma_{\text{lo}} (\gamma_{\text{lo}} - \gamma_{\text{hi}}) - \me^{-2\tau_{\text{env}}/\tau}(\gamma_{\text{lo}} - \gamma_{\text{hi}})^2 + 2 \gamma_{\text{lo}} \gamma_{\text{hi}} + \gamma_{\text{hi}}^2 \r + 2 \gamma_{\text{lo}}^2 \nn \\
&& B = \f{1}{(\gamma_{\text{lo}}-2)\gamma_{\text{lo}}} \l -2(\gamma_{\text{lo}}-1)^2 + \f{\tau}{\tau_{\text{env}}} \l (\gamma_{\text{lo}}-2) \ln \f{\gamma(t)}{\gamma_{\text{hi}}} + \gamma_{\text{lo}} \ln \l \f{2-\gamma_{\text{hi}}}{2-\gamma(t)} \r   \r \r
\eea

\noindent plotted in Fig. 4c of the main text. Note that Eq. \ref{eq:somethingofamouthful} is a function only of the ratio $\tau/\tau_{\text{env}}$.

\bibliographystyle{unsrt}